# Intrinsic and extrinsic defect-related excitons in TMDCs

*Kyrylo Greben[1,\*], Sonakshi Arora[1,2], Moshe G. Harats[1], and Kirill I. Bolotin[1,\*]*

[1] Department of Physics, Freie Universität Berlin, 14195 Berlin, Germany

[2] Department of Quantum Nanoscience, Faculty of Applied Science, Delft University of Technology, 2628 CJ, Delft, The Netherland

*k.greben@fu-berlin.de

*kirill.bolotin@fu-berlin.de

**ABSTRACT**

We investigate an excitonic peak appearing in low-temperature photoluminescence of monolayer transition metal dichalcogenides (TMDCs), which is commonly associated with defects and disorder. First, to uncover the intrinsic origin of defect-related excitons, we study their dependence on gate voltage, excitation power, and temperature in a prototypical TMDC monolayer, $MoS_2$. We show that the entire range of behaviors of defect-related excitons can be understood in terms of a simple model, where neutral excitons are bound to ionized donor levels, likely related to sulphur vacancies, with a density of $7\times10^{11}$ cm$^{-2}$. Second, to study the extrinsic origin of defect-related excitons, we controllably deposit oxygen molecules *in-situ* onto the surface of $MoS_2$ kept at cryogenic temperature. We find that in addition to trivial p-doping of $3\times10^{12}$ cm$^{-2}$, oxygen affects the formation of defect-related excitons by functionalizing the vacancy. Combined, our results uncover the origin of defect-related excitons, suggest a simple and conclusive approach to track the functionalization of TMDCs, benchmark device quality, and pave the way towards exciton engineering in hybrid organic-inorganic TMDC devices.

## INTRODUCTION

Monolayer Transition Metal Dichalcogenides (TMDCs) are semiconductors with a direct bandgap in the visible range[1,2]. Due to the weak screening and strong electron-hole interactions in these materials, their optical properties are dominated by excitons, bound electron-hole pairs.

To date, the properties of free neutral excitons and exciton complexes such as charged excitons (trions) have been studied and largely understood[3,4]. Binding energies[5–7], formation and dissociation mechanisms[8,9], coherence effects[10], and spin-valley effects[11–17] of these excitons have been identified. In addition to neutral and charged excitonic peaks, a feature that is often attributed to localized, rather than free excitons appears in photoluminescence (PL) spectra of many TMDC devices at low temperatures[18–26]. While it is widely assumed that this feature is related to defects in TMDCs, multiple questions remain unanswered.

First, what is the nature of the defect-related PL feature (D peak)? It has been attributed both to two- and three-particle states as well as to various defect types[18,20,21,27–30]. Second, can the D peak be used as an indicator of a sample quality, i.e. to determine the concentration and type of defects? Third, are defect-related excitons of an extrinsic or intrinsic origin? Previously, the D peak has been ascribed separately to intrinsic structural defects[28,30,31] in TMDCs or to extrinsic impurities on TMDC surface[32–35]. The final and the most interesting question is whether the D peak can be used to gauge chemical functionalization of TMDCs. The field of TMDC functionalization has been quickly developing in recent years thanks to potential applications of functionalized materials as chemical- and bio-sensors[36–39]. Defects are centers of chemical activity in otherwise inert TMDCs, and hence are critical for functionalization[40–45]. At the same time, simple techniques to determine the success and extent of functionalization are lacking[46,47]. Here, we address these questions and elucidate the nature of D excitons.

We study the evolution of the PL spectra of a monolayer $MoS_2$ in the region of the defect-related peak while tuning multiple experimental variables. In addition to well-studied temperature and laser excitation power dependencies, we analyze the behavior of the D peak with gate voltage and surface functionalization due to oxygen molecules. We then show that all our observations can be understood within a simple mass-action law type model[44,48]. In this model, the defect-related exciton is described as a neutral exciton bound to ionized donor levels close to the edge of the conduction band of $MoS_2$. These levels are likely related to sulphur vacancies[28,29]. Finally, by controllably depositing oxygen molecules onto $MoS_2$ *in-situ* at cryogenic temperatures, we show that the D peak additionally has an extrinsic character, i.e. is influenced by the impurities on the surface of $MoS_2$. We show that this influence is likely caused by oxygen functionalization of sulfur vacancies.

## RESULTS

### Setup

Our custom setup is designed both for gated low-temperature PL measurements as well as for measurements involving *in-situ* sample annealing and deposition of oxygen molecules. (Fig. 1a). The samples are studied inside an optical cryostat at temperatures ranging from 7 to 300 K. For optical characterization, we use Nikon 50x SLWD objective with NA = 0.5 and the laser excitation wavelength of 532 nm, with power between 0.15 and 30 µW. For gate-dependent PL measurements, we use monolayer $MoS_2$ field effect transistors (FETs) (Fig. 1b, Inset), fabricated on 300 nm $SiO_2$/p-Si substrate.

In general, two complications can arise in any experiment attempting controlled deposition of molecules onto a TMDC. First, the surface of an as-prepared TMDC is always covered by a layer of contaminants (e.g. water and organic molecules). Second, deposited molecules may diffuse and merge into clusters on the surface. We developed an approach tackling these difficulties. To remove the layer of contaminants, we microfabricated a Cr/Au (3 nm/70 nm) heater and a thermometer on the same chip in proximity to our sample (Fig.1b). This allows *in-situ* high-temperature annealing (> 400 K) within a few seconds, while the rest of the cryostat is kept at a base temperature (T = 7 K). To avoid problems associated with molecule diffusion and clustering, we added a feed-through nozzle with a 100 µm diameter aperture to our cryostat (Fig. 1a). Oxygen is deposited *in-situ* through this nozzle on top of the sample kept at the base temperature, thus promoting the sticking of oxygen molecules to the surface of $MoS_2$.

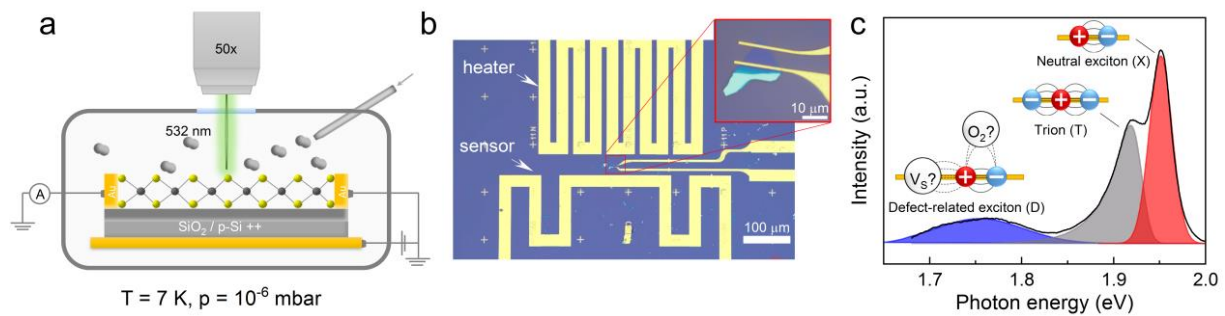

**Figure 1: Experimental setup and a typical measurement.** (**a**) A cryogenic PL setup customized for in-situ annealing and gas deposition. (**b**) An optical image of the device showing an electrically contacted $MoS_2$ flake as well as an on-chip resistive thermometer and heater. (**c**) Typical PL spectrum of $MoS_2$ at cryogenic temperatures showing peaks due to neutral excitons, trions, and defect-related excitons.

## Evolution of the low-temperature PL spectra of MoS₂

We study defect-related excitons using low-temperature PL spectroscopy. From the family of TMDCs, we have chosen $MoS_2$ as a prototypical material with a significant density of intrinsic defects[18,28,29]. A typical PL spectrum of $MoS_2$ at cryogenic temperatures exhibits three peaks (Fig. 1c). The peak at 1.96 eV, labelled "X", is associated with neutral excitons, while the peak at 1.93 eV, labelled "T" is related to negatively charged excitons (trions). In addition, the peak at 1.77 eV appears in many (but not all) measured samples. This peak, labelled "D", is assumed to relate to disorder[21–23,49–53]. It is the main subject of this work.

In order to elucidate the nature of defect-related excitons in $MoS_2$, we systematically tune three experimental variables: temperature $T$, laser excitation power $P$, and the position of the Fermi level $E_F$ controlled by the backgate voltage $V_G$ (Fig. 2a). We notice that the area under the D peak increases for negative gate voltages, low temperatures, and low illumination powers (Fig. 2b). Conversely, it decreases at high positive gate voltages and completely disappears (at any $V_G$) for temperatures above 240 K (Fig. 2b). Overall, the D peak changes by over *three* orders of magnitude with $V_G$ and over *four* orders of magnitude with $T$.

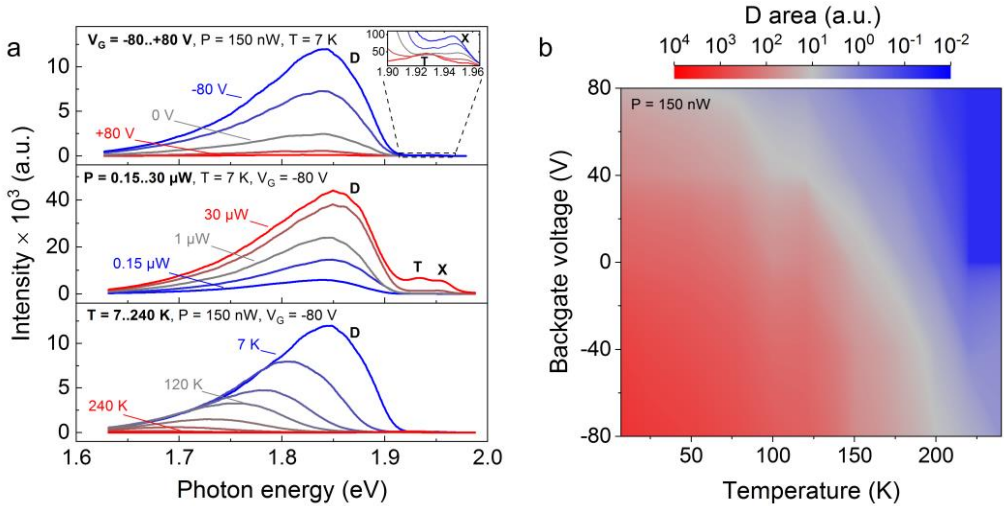

**Figure 2: Gate ($V_G$), temperature (T), excitation power (P) dependence of monolayer MoS₂ photoluminescence.** (**a**) PL spectra of monolayer $MoS_2$ as a function of $V_G$ (top), P (middle) and T (bottom). The inset shows the region of neutral and charged excitons. (**b**) Color map showing $V_G$ and T dependence of the area under the D peak.

## Defect-related exciton analysis

Different physical models for defect-related excitons are expected to produce different $V_G$-, $P$- and $T$-dependencies of X, T, and D peaks. We will now show that the dependencies we observe (Fig. 2) are consistent with a single model: neutral excitons bound to ionized shallow donor levels. Such levels may originate from sulphur vacancies. We start by developing a simple description of our data that is based on mass action law type equations. In this description, the co-existence of neutral excitons (X), trions (T) and free electrons (*e*) can be viewed as a chemical reaction that has reached its equilibrium, $X + e \leftrightarrow T$ (Supplementary Information and Ref. 48). Similarly, the formation and dissociation of a defect-related exciton (D) from an ionized donor (*d*), and a neutral exciton can be viewed as a reaction $X + d \leftrightarrow D$. The equality of the rates of forward and reverse reactions leads to the following equations:

$$n_T/n_X = K_T \cdot n_e \qquad (1)$$
$$n_D/n_X = K_D \cdot N_D \qquad (2)$$

Here, $n_X$, $n_T$, $n_e$, are the concentrations of neutral excitons, trions, and of free electrons respectively, while $n_D$ and $N_D$ are concentrations of defect-related excitons and *unoccupied ionized* defect levels. The rate constants $K_T \sim T \cdot exp(-E_T/k_B T)$ and $K_D \sim T \cdot exp(-E_D/k_B T)$ are related to the trion binding energy $E_T$ and the binding energy of the defect-bound exciton $E_D$. While Eq. 1 was used before to describe exciton/trion equilibrium in TMDCs[44,48], Eq. 2 has not been previously considered, to the best of our knowledge. Both equations can be viewed as a limiting case of a more complex system of equations describing more processes (e.g. creation of free electrons from exciton recombination) in the limit of low excitation powers and long defect exciton lifetime[54].

Equations 1 and 2 provide a simple approach to directly extract the carrier density $n_e$ and the concentration of ionized levels $N_D$ from our experimental data. We fit the spectra in Fig. 2a using Gaussian peaks for defect-bound and neutral excitons. For trions, we use and a more complex fitting function taking into account the electron recoil effect[62] (see Fig. 1c). We then extract the spectral areas under the trion, defect-related, and neutral exciton peaks, $A_T$, $A_D$, and $A_X$, and assume that they are proportional to the respective exciton densities[44]. We extract the experimental dependencies of area ratios $A_T/A_X$ and $A_D/A_X$ on $V_G$, $P$, and $T$ from the data in Fig. 2, and plot the resulting analysis in Fig. 3. Finally, from Eqs. 1 and 2 we get $n_e \sim n_T/n_X \sim A_T/A_X$ and $N_D \sim n_D/n_X \sim A_D/A_X$.

We will now show that complex $V_G$-, $P$-, and $T$- dependencies of $n_e \sim A_T/A_X$ and $N_D \sim A_D/A_X$ seen in Fig. 3 have a simple explanation assuming that the density of electrons is controlled by gating and the D peak corresponds to neutral excitons bound to ionized donor levels.

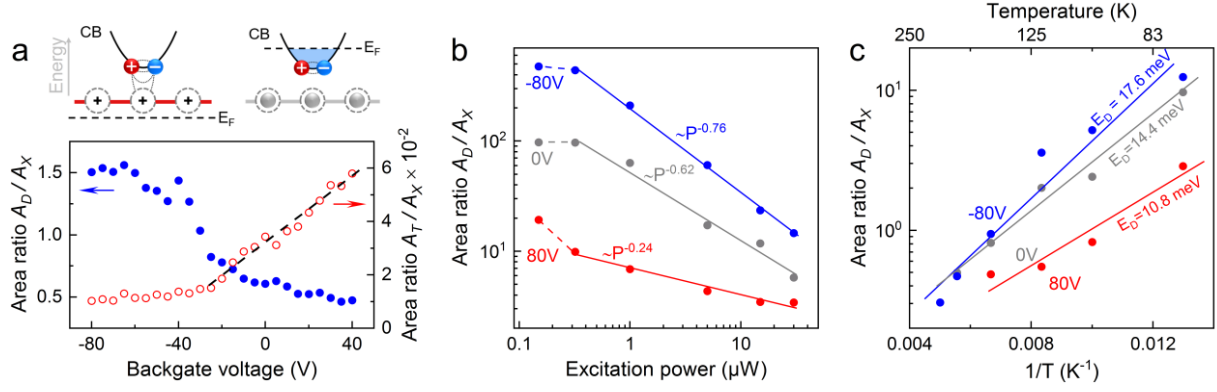

**Figure 3: Analysis of defect related excitons.** (**a**) $V_G$-dependence of $A_D/A_X$, the ratio of areas between defect-related and neutral excitonic peaks (full blue circles, left axis), and $A_T/A_X$, the ratio of areas between charged and neutral excitonic peaks (open red circles, right axis). The black dashed line is a linear fit. The insets depict the model for defect-bound excitons proposed here: a neutral exciton bound to an ionized donor level. These levels are only ionized when the Fermi level is below these donor levels. (**b**) $A_D/A_X$ vs. excitation power at T = 7 K. Data at different $V_G$ is shown in different colors. Full circles are the experimental points, the solid lines are the power-law fits. (**c**) Arrhenius plot of $A_D/A_X$ for P = 150 nW at different $V_G$ shown in different colors. Full circles are experimental points and the solid lines are linear fits.

The $V_G$-dependence of both $A_T/A_X$ and $A_D/A_X$ is plotted in Fig. 3a. This data is taken at $T$ = 130 K and $P$ = 1 µW to ensure that the defect-related peak does not dominate the spectrum and that Eq. 1 applies (Fig. 2a and Supplementary Information). We see that $A_T/A_X$ is small and approximately constant below $V_G \approx$ -25 V while increasing roughly linearly with $V_G$ above this value (Fig. 3b, open red circles). This is exactly the behavior expected for $n_e$: it is zero when $E_F$ is inside the bandgap and $n_e \sim C_G V_G$, where $C_G$ is the gate capacitance, when $E_F$ is above the conduction band minimum (Supplementary Fig. S3). This suggests that the conduction band minima are located at about $V_G \approx$ -25 V (Supplementary Information). At the same time, $A_D/A_X$ increases with decreasing $V_G$ (Fig. 3a, full blue circles). The behavior of $A_D/A_X$ matches the expected $V_G$-dependence of $N_D$ for a specific defect type: ionized donor level. For large negative $V_G$, $E_F$ is deep inside the bandgap, and all impurity levels are ionized. In this case, $N_D$ is simply the defect density. In contrast, large positive $V_G$ corresponds to $E_F$ in the conduction band (Fig. 2a, inset). In this case, most of the impurity levels are neutral and hence are unavailable for exciton binding. The rapid increase of $A_D/A_X$ close to the point of $E_F$ crossing into the band edge, $V_G \approx$ -25 V, suggests that the defects are relatively shallow ionized donors. This defect type is consistent with sulphur vacancies which lie 0.3—0.7 eV below the conduction band minimum according to DFT calculations[20,29,52].

Next, we analyze the behavior of $A_D/A_X$ and $A_T/A_X$ with $P$ (Fig. 3b and Supplementary Fig. S1a). $A_T/A_X$ is constant at low powers while growing for $P > 1$ µW. This is also the behavior expected for $n_e$. At small excitation powers, $n_e$ is equal to the density of background carriers in the sample, while at higher $P$ additional carriers are generated by recombination processes[48]. In contrast, $A_D/A_X$ decreases with $P$ roughly following the power law $A_D/A_X \sim P^a$ with the exponent ranging from $a \sim 0.7$ at $V_G$ = -80 V to $a \sim 0.3$ at $V_G$ = 80 V. In addition, a saturation region is observed at large negative $V_G$ at the smallest $P < 0.5$ µW. This behavior is also consistent with that expected for $N_D$. For low excitation powers, ionized defect sites become bound by excitons and $N_D$ is close to the defect density. As $P$ increases, there are more excitons generated than the defect sites available, leading to an overall drop in $N_D \sim A_D/A_X$ (see also Supplementary Fig. S5). One can show that under our experimental conditions, the illumination power at which the crossover between the two regimes occurs is related to the defect density (Supplementary Information). From the experimentally-observed crossover at around 0.5 µW, we estimate $N_D = 7 \times 10^{11} cm^{-2}$, close to the density of the sulphur vacancies obtained by other experimental methods[20,30]. Evolution of the saturation point with $V_G$, evident in Fig. 3b, is also expected in our model, since $N_D$ decreases with increasing $V_G$ (see Fig. 3a). The power law dependence seen in Fig. 3b is consistent with results of calculations[54] and experimental observations[20,26,55].

Finally, we discuss the temperature dependence of $A_D/A_X$ and $A_T/A_X$ (Fig. 3c and Supplementary Fig. 1b). $A_T/A_X$ is weakly temperature-dependent and the trion peak is still visible at room temperature. This is consistent with temperature-independent $n_e$ and large trion binding energy entering into the rate constant $K_T$[5]. In contrast, $A_D/A_X$ strongly decreases with temperature, disappearing for temperatures above 240 K (Fig. 3c). One possible source for the temperature dependence of $A_D/A_X$ is the rate constant $K_D$ in Eq. 2. The activation energy, $E_D$, entering into it, should relate to the energy difference between neutral and defect-related excitons, $\approx 150$ meV. However, a much smaller activation energy, $\approx 17.6$ meV at $V_G$ = -80 V, is extracted from Fig. 3c, in agreement with earlier observations[25]. Therefore, we believe that $N_D$ is the dominant source of the $T$-dependence of $A_D/A_X$. Sulphur vacancies have been calculated to produce a nearly degenerate doublet with an energy spacing of an order of $\approx 14$ meV[51]. We speculate that the redistribution of carriers between these levels contributes to the $T$-dependence seen in Fig. 3c.

Overall, we see that the dependence of the defect-bound exciton peak on all experimental variables is consistent with that of neutral exciton bound to an ionized donor level likely related to sulphur vacancies. To the best of our knowledge, no other model for defect-related excitons reported in the literature fits the data of Figs. 2 and 3. First, the carrier density-dependent screening of defect-bound excitons can potentially explain the data similar to Fig. 3a. However, the screening is only effective at carrier densities of order $a^{-2}$, where $a$ is the real-space exciton size[56,57]. Available estimates put this density at $10^{13}$ cm$^{-2}$. In contrast, the largest effect is seen in our data when the Fermi level is inside the bandgap and the density of delocalized electrons is zero. Second, various acceptor levels may be present in MoS$_2$[25]. However, such defects could not explain the data of Fig. 3a, where the defect peak appears immediately after $E_F$ crosses from the conduction band into the bandgap. Finally, it has been suggested that the defect-related PL peak may result from a two-particle state related to recombination between a free hole and a neutral donor impurity[27]. However, such a state should be favored when $E_F$ is in the conduction band, opposite to what we observe in Fig. 3a.

**Extrinsic contribution**

Our data so far suggests that the defect-related PL feature stems from neutral excitons bound to an ionized donor level. Such a level may be intrinsic, e.g. originate from a lattice defect in TMDCs such as sulphur vacancies[28,30,31]. However, the defect level may also be extrinsic, and originate from an impurity molecule[32–34]. DFT calculations do suggest that shallow states are affected by adsorbed organic molecules and gases[51,53,58–60]. To further understand the extrinsic vs. intrinsic character of the defect level, we controllably deposited a common molecule, oxygen[34,58], onto the surface of MoS$_2$ kept at cryogenic temperature and examined the evolution of the defect-related peak, D, with time-resolution.

We observe that the PL spectra change dramatically due to annealing and oxygen deposition, especially in the region of the D peak (Fig. 4a). To quantify these changes, we plot $A_D/A_X$ and $A_T/A_X$ at a constant $V_G = 0$ V vs. time with 1 min steps during the deposition process (Fig. 4b). During the first 15 minutes, we observe a drop of $A_T/A_X$ and a rise of $A_D/A_X$ followed by saturation in both quantities. We note that the normalization of $A_T$ and $A_D$ by the neutral exciton area, $A_X$, accounts for possible changes in PL due to the transparency of deposited layers of molecules. Using the data presented in Fig. 3a,b, we suggest a simple explanation to these trends. Given that our MoS$_2$ is n-doped, time-dependent behavior of both $A_T/A_X$ and $A_D/A_X$ is consistent with $E_F$ decreasing and moving into the bandgap (Fig. 3a). That, in turn, is indicative of the charge transfer from the TMDC to O$_2$ molecules. Such charge transfer has been previously seen experimentally[34,61] and predicted computationally[58]. The saturation of charge transfer after 15 minutes suggests that the interaction between the oxygen molecules and a TMDC becomes negligible after full surface coverage. Two questions remain, however. What is the total density of the transferred charge, and can we explain the spectral changes observed in Fig. 4a simply by a charge transfer model?

To address these questions, we compare the $V_G$-dependence of the PL spectra before and after $O_2$ deposition. We pick the spectra of as-exfoliated, annealed, and functionalized states of the sample for which $A_T/A_X$ is the same. We find that the spectra with $V_G = -60$ V for as-exfoliated, $V_G = -40$ V for annealed and $V_G = 0$ V for the spectrum after $O_2$ deposition have equal $A_T/A_X$ ratios (Fig. 4c). As matching of $A_T/A_X$ indicates equal carrier densities, we conclude that additional carriers produced by the field effect exactly compensate for charge transfer due to the presence of molecules. Therefore, we can calculate the amount of charge transfer in each case using the relation $\Delta n = C_G \Delta V_G$, where $C_G = 7.8 \times 10^{10} V^{-1} cm^{-2}$ is the gate capacitance. We obtain a carrier density of $\approx 2.9 \times 10^{12} cm^{-2}$ due to saturated $O_2$ deposition. This is close to the density of the full surface coverage obtained from DFT calculations[34,58]. The removal of adsorbates from the surface of $MoS_2$ extracts $1.4 \times 10^{12} cm^{-2}$ carriers. Therefore, the adsorbates are *n*-dopants.

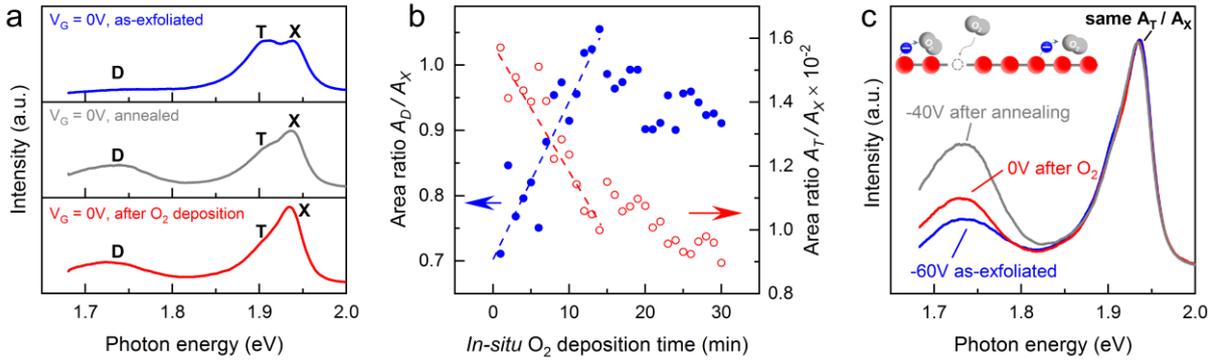

**Figure 4: The extrinsic nature of the defect-related excitonic peak.** (**a**) PL spectra of $MoS_2$ at $V_G = 0$ V at the beginning of measurements (top) after in-situ thermal annealing (middle) and after in-situ $O_2$ deposition (bottom). (**b**) Changes in $A_D/A_X$ (full blue circles, left axis) and $A_T/A_X$ (open red circles, right axis) with time during in-situ $O_2$ deposition. (**c**) PL spectra of as-exfoliated at $V_G = -60$ V, after annealing at $V_G = -40$ V, and after $O_2$ deposition at $V_G = 0$ V. Same $A_T/A_X$ indicates that all three curves correspond to same carrier density in $MoS_2$.

The most interesting feature of Fig. 4c is that although all curves are matched in the region of the neutral and charged excitons, they starkly differ in the region of the D peak. Specifically, the concentration of unoccupied defect sites $N_D \sim A_D/A_X$ increases after annealing and then drops after $O_2$ deposition. Therefore, our data cannot be simply explained as a result of charge transfer. Instead, this observation indicates that the defect-related peak has at least partially extrinsic character. We speculate that defect-related excitons interact with oxygen molecules on the surface of $MoS_2$ through passivation of a sulfur vacancy by an oxygen molecule, known to eliminate the midgap states accessible for excitons[35]. The increase of $N_D$ after annealing and its subsequent drop after functionalization is consistent with removing and then depositing molecules.

In conclusion, we investigated the dependence of the defect-related feature in the PL spectrum of $MoS_2$ on multiple experimental variables: temperature, excitation power, gate voltage, and surface coverage. Our data is consistent with a single model for the defect-related exciton: a neutral exciton bound to a shallow ionized donor level close to conduction band minimum. This level likely originates from a sulphur vacancy, but is influenced by oxygen passivation of the defect. To reveal the extrinsic contribution to the defect-related excitons, we developed an approach distinguishing the effects of doping from that of excitons interacting with the defects. Our results have several interesting implications. First, our data allows discriminating between multiple models for the defect-related excitons discussed in the literature. It is inconsistent with the models involving acceptor levels or valence band-midgap state transitions. Second, we show that the presence and the height of the D peak cannot be directly used as an indicator of sample quality, despite the appeal of that simple metric. Instead, a comparison between the samples with the same Fermi level is required. Finally, we prove that molecules on the surface of a TDMC influence midgap states of the TMDC. On one hand, this highlights the necessity of pristine device to study excitonic physics in TMDCs. On the other hand, surface functionalization may open interesting avenues towards controlled defect-engineering of excitonic properties.

## ACKNOWLEDGEMENTS

We gratefully acknowledge Denis Yagodkin and Benjamin I. Weintrub for useful discussions and comments. This work was supported by the Deutsche Forschungsgemeinschaft (DFG) - Projektnummer 182087777 - SFB 951 and ERC Starting grant no. 639739.

# Intrinsic and extrinsic defect-related excitons in TMDCs

*Kyrylo Greben[1,*], Sonakshi Arora[1,2], Moshe G. Harats[1], and Kirill I. Bolotin[1,*]*

[1] Department of Physics, Freie Universität Berlin, 14195 Berlin, Germany
[2] Department of Quantum Nanoscience, Faculty of Applied Science, Delft University of Technology, 2628 CJ, Delft, The Netherland

*k.greben@fu-berlin.de
*kirill.bolotin@fu-berlin.de

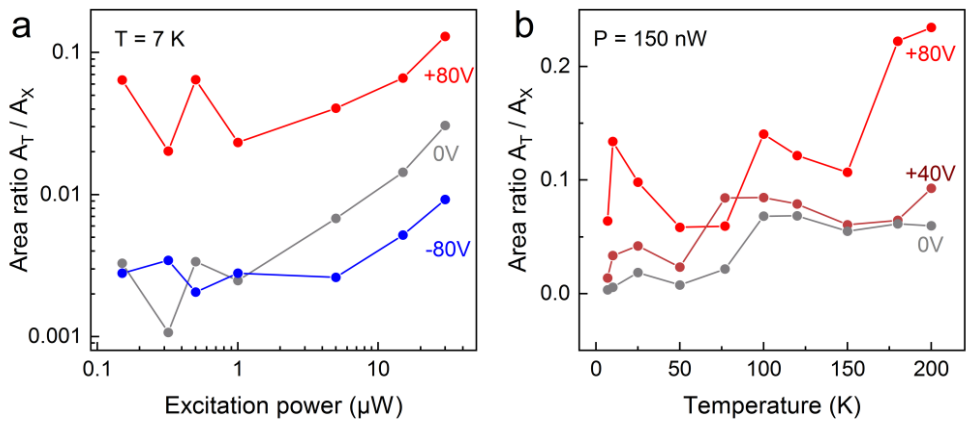

**Supplementary Figure S1: Temperature and excitation power dependence of the $A_T/A_X$ ratio.** (**a**) The ratio between spectral areas below the trion peak ($A_T$) and neutral exciton peak ($A_X$) of monolayer $MoS_2$ plotted vs. excitation power. Curves for different backgate voltages ($V_G$) are shown in different colors. A saturation region below 1 µW is visible. (**b**) $A_T/A_X$ vs. temperature, for several different $V_G$.

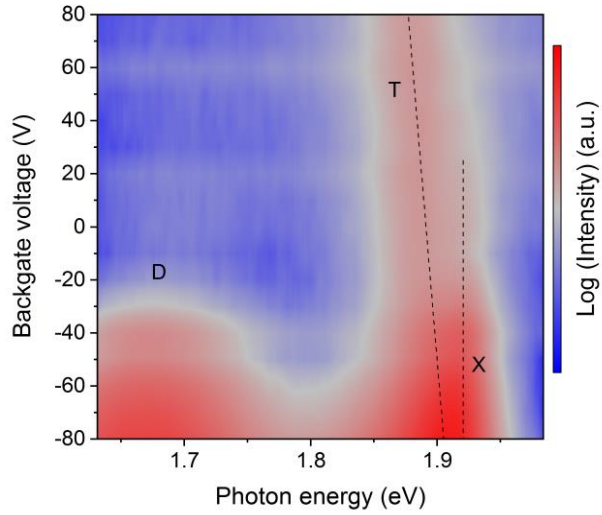

**Supplementary Figure S2: Extracting Fermi level position from spectral positions of neutral and charged excitons.** PL spectra of monolayer $MoS_2$ vs. backgate voltage ($V_G$), plotted as color map. Distinct features corresponding to neutral (X), charged (T) and defect-related (D) excitons are marked. The spectral positions of X and T are denoted with dashed lines. Merging of X and T peaks for $V_G < -40$ is consistent with the $E_F$ reaching the bottom of the conduction band. We use the analysis of Ref. S1 to extract gate dependence of the Fermi level ($E_F$). Quantitatively, $E^X - E^T = E_F + E_T$, where $E_T \approx 22$ meV is the trion binding energy. The numerical value of the slope $d[E^X(V_G) - E^T(V_G)]/dV_G \sim 0.166$ meV/V corresponds to the effective mass, $m^* = 0.52 m_0$, close to numerically calculated effective mass for $MoS_2$, $0.35 m_0$ (Ref.S2).

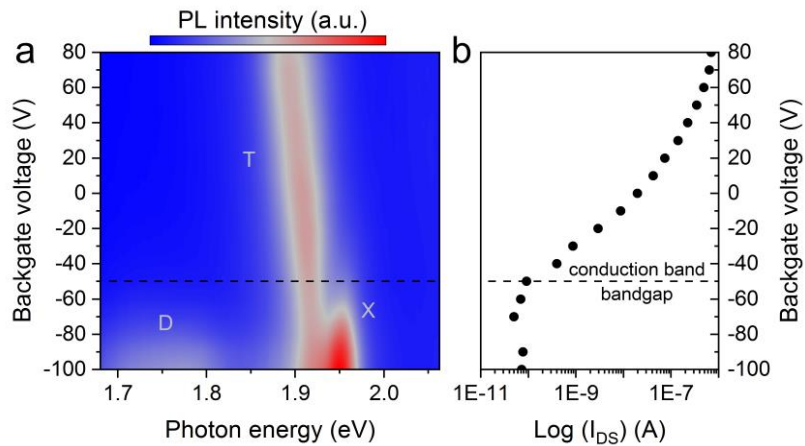

**Supplementary Figure S3: Probing defect level position using gate-dependent photoluminescence and electrical transport measurements.** (a) PL spectra vs. backgate

voltage $V_G$, plotted as color map for a different sample B. Features related to neutral (X), charged (T) and defect-related (D) excitons emission are marked. The D peak appears only for $V_G < -50$ V. (**b**) The drain-source current ($I_{DS}$) vs. $V_G$ for the sample is measured in a field-effect transistor (FET) geometry. At $V_G < -50$ V the FET switches off. This indicates that $E_F$ reaches the bottom of the conduction band at this voltage. The data of (a) and (b), taken together, suggest that defect-related levels lie inside the bandgap, close to the top of the conduction band.

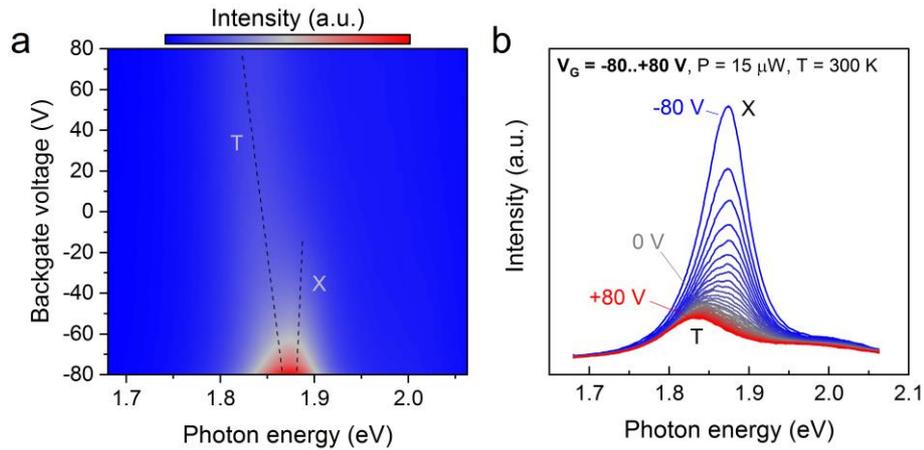

**Supplementary Figure S4: Backgate-dependent PL spectra of MoS$_2$ at room temperature.** (**a**) A color map of PL spectra, showing the backgate dependence of neutral (X) and charged (T) exciton features. The dashed lines indicate the spectral positions of X and T. (**b**) Corresponding individual PL spectra. Different colors correspond to different backgate voltages.

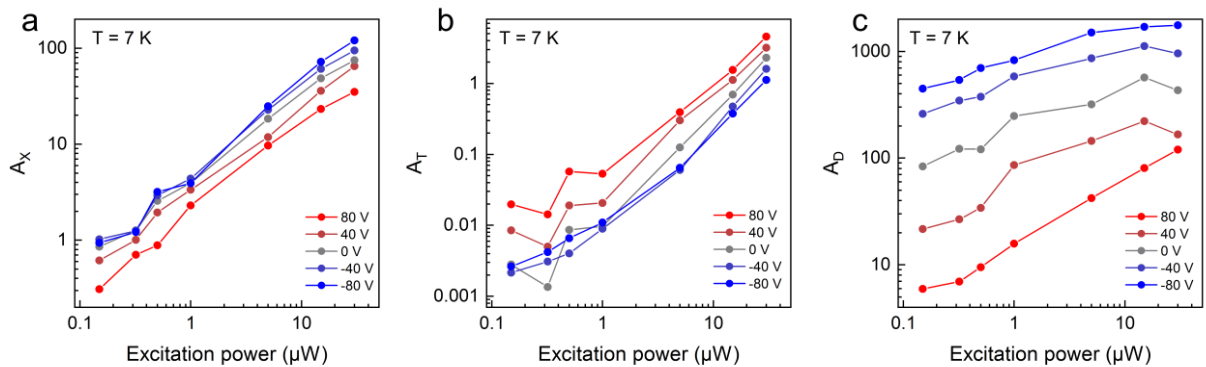

**Supplementary Figure S5: The laser excitation power dependence of the spectral areas under neutral (a), charged (b) and defect-related (c) excitons.** Different colors indicate different backgate voltages. The data is taken at the temperature of 7 K.

**Trion/neutral exciton analysis**

The behavior of the neutral exciton and trion can be understood within a simple rate-equation model, as has been first shown in Supplementary Ref. S3. The co-existence of neutral excitons (X), trions (T), and free electrons ($e$) can be viewed as a chemical reaction that has reached its equilibrium. The equality of the rates of forward and reverse reactions $X + e \leftrightarrow T$ leads to the mass-action law type equation:

$$n_T/n_X = K_T \cdot n_e \quad (S1)$$

Here $n_X$, $n_T$, $n_e$ are the concentrations of neutral excitons, trions, and of free electrons respectively, and $K_T \sim T \cdot exp(-E_T/k_B T)$ is the rate equation constant related to the trion binding energy $E_T$.

We use Supplementary Eq. S1 to analyze the spectra presented in Fig. 2 of main text. To accomplish this, we obtain the areas and spectral positions of excitonic peaks through fitting (see Main text). We then extract the spectral areas under the trion and neutral exciton peaks, $A_T$ and $A_X$, and assume that they are proportional to the respective exciton densities[S4]. From Eq. 1 we get $A_T/A_X \sim n_T/n_X \sim n_e$. It is important to note that in general $n_e$ in this equation is the sum of the background electron density, $n_b$, and the density of photo-excited carriers that is excitation power dependent. However, when the illumination intensity is small enough, $n_e \sim n_b$ (Supplementary Fig. S1a) and the analysis of the Eq. S1 is especially simple.

In the Supplementary Fig. S2 and Fig.3a from the main text, we see that the neutral and defect-related excitons dominate the PL below $V_G \approx$ -25 V, leading to a small and approximately constant $A_T/A_X$. From the combination of electrical and optical measurements in Supplementary Fig. S3 we know, that trions start to dominate the PL spectrum as soon as the Fermi level ($E_F$) enters the conduction band. This is due to the $V_G$-dependent behavior of $n_e$: it should depend on $V_G$ as $n_e \sim C_G V_G$, where $C_G$ is the gate capacitance when $E_F$ is above the conduction band minima, and $n_e = 0$ when $E_F$ is inside the bandgap. In Supplementary Fig. S2 and Fig. 3a the conduction band minima is located nearly at $V_G \approx$ -25 V. Above this value, $A_T/A_X$ increases roughly linearly with $V_G$. The position of $E_F$ relative to the conduction band minima is further confirmed from the energy separation between the X and T peaks, indicated with dashed lines in Supplementary Fig. S2. The numerical value of the slope d[$E^X(V_G)$-$E^T$ $(V_G)$]/d$V_G$ ≈ 0.166 meV/V corresponds to the effective mass, m* = 0.52m$_0$, close to numerically calculated effective mass for MoS$_2$, 0.35m$_0$ (Ref.S2).

We also note that $A_T/A_X$ has relatively weak T-dependence (Supplementary Fig. S1b), indicating that trions survive up to room temperature as was previously shown in Supplementary Refs. S3 and S5 and is confirmed in Supplementary Fig. S4. This relatively weak temperature dependence is consistent with the binding energy of the trion, $E_T \sim k_B T \sim 25\ meV$, entering the rate constant, $K_T$. Finally, both $A_T$ and $A_X$ depend linearly on the excitation power (Supplementary Fig. S5), as expected for free excitons[S6].

**Estimating the concentration of defects from saturation measurements**

The concentration of defects can be estimated from the experimentally observed saturation in $A_D/A_X$ excitation power dependence (Fig. 3b). Indeed, such saturation is expected when every photoexcited carrier binds to an empty defect level within a lifetime that level (the lifetime of defect-bound exciton). Assuming that all photoexcited excitons eventually bind to defects if they are available, we obtain the following simple estimate for the defect concentration $n_d$:

$$n_D = \frac{C}{A} \times \frac{P_{sat}}{E_{ph}} \times t \ \ (S1),$$

Where $C \sim 7\%$ is the absorption coefficient in MoS$_2$[S5]; $A = \frac{\pi}{4} \times (\frac{1.22\lambda}{NA})^2$ is the area of the laser illumination spot (NA = 0.5 is a numerical aperture of our objective and $\lambda$ = 532 nm is the excitation wavelength); $P_{sat}$ is the laser excitation power corresponding to the on-set of saturation, and $t \approx 100$ ps is the defect-related exciton lifetime[S7–9]. From this equation we obtain $n_D = 7 \times 10^{11}$ cm$^{-2}$ for the experimentally observed onset of saturation at $P_{sat}$ = 500 nW.

**SUPPLEMENTARY REFERENCES**

S1. Chernikov, A. *et al.* Electrical Tuning of Exciton Binding Energies in Monolayer WS2. *Phys. Rev. Lett.* **115**, 1–6 (2015).

S2. Cheiwchanchamnangij, T. & Lambrecht, W. R. L. Quasiparticle band structure calculation of monolayer, bilayer, and bulk MoS2. *Phys. Rev. B* **85**, 205302 (2012).

S3. Ross, J. S. *et al.* Electrical control of neutral and charged excitons in a monolayer